\documentclass[conference]{IEEEtran}[12pt]
\usepackage{hyperref}	
\usepackage{setspace}
\usepackage{rotating}

\usepackage{makeidx}  
\usepackage{url}
\usepackage{latexsym}
\usepackage{graphicx,paralist,amsmath,mathrsfs,multirow,amssymb,tabularx,color}
\usepackage{subcaption}
\usepackage[ruled,longend]{algorithm2e}
\usepackage{soul}
\usepackage{caption}
\usepackage{fontawesome}
\usepackage[stable]{footmisc}
\usepackage{float}

\ifCLASSINFOpdf
\else
\fi
\usepackage{array}
\newcolumntype{P}[1]{>{\centering\arraybackslash}p{#1}}
\newcolumntype{M}[1]{>{\centering\arraybackslash}m{#1}}
\usepackage{fancyhdr}
\usepackage{kantlipsum}
\fancypagestyle{plain}{
\fancyhf{}
\fancyhead[C]{Conference on \LaTeX} 

}
\usepackage{eso-pic}
\IEEEoverridecommandlockouts
\makeatletter
\def\@IEEEpubidpullup{8\baselineskip}
\makeatother

\usepackage{fancyhdr}
\usepackage{kantlipsum}
\fancypagestyle{plain}{
\fancyhf{}
\fancyhead[C]{Conference on \LaTeX} 

}
\usepackage{eso-pic}

\begin{document}

\AddToShipoutPictureBG*{
\AtPageUpperLeft{
\setlength\unitlength{1in}
\hspace*{\dimexpr0.5\paperwidth\relax}
\makebox(0,-0.75)[c]{\textbf{2020 IEEE/ACM International Conference on Advances in Social Networks Analysis and Mining (ASONAM)}}}}


%

\title{Detecting Fake News Spreaders in Social Networks using Inductive Representation Learning}

\author{Bhavtosh Rath, Aadesh Salecha, Jaideep Srivastava \\
Dept. of Computer Science \& Engineering, University of Minnesota, Twin Cities, MN, USA\\
rathx082@umn.edu, salec006@umn.edu, srivasta@umn.edu\\
}


%



\maketitle

\IEEEoverridecommandlockouts
\IEEEpubid{\parbox{\columnwidth}{\vspace{8pt}
\makebox[\columnwidth][t]{IEEE/ACM ASONAM 2020, December 7-10, 2020}
\makebox[\columnwidth][t]{978-1-7281-1056-1/20/\$31.00~\copyright\space2020 IEEE} \hfill} \hspace{\columnsep}\makebox[\columnwidth]{}}
\IEEEpubidadjcol

\begin{abstract}
An important aspect of preventing fake news dissemination is to proactively detect the likelihood of its spreading. Research in the domain of fake news spreader detection has not been explored much from a network analysis perspective. In this paper, we propose a graph neural network based approach to identify nodes that are likely to become spreaders of false information. Using the community health assessment model and interpersonal trust we propose an inductive representation learning framework to predict nodes of densely-connected community structures that are most likely to spread fake news, thus making the entire community vulnerable to the infection. Using topology and interaction based trust properties of nodes in real-world Twitter networks, we are able to predict false information spreaders with an accuracy of over 90\%.
\end{abstract}


%
\IEEEpeerreviewmaketitle

\section{Introduction}
 
People use social networking platforms like Twitter, Facebook and Whatsapp not only to consume and share information but also their opinions about it. Ease of sharing has made it possible to spread information quickly, often without verifying it, resulting in fake news spreading. This has led to increase in interest among social media researchers to propose  fake news spreading detection models. In this context, it is not only important to detect false information but also identify people who are  most likely to believe and spread the false information. This is so because detection of fake news spreaders can help contain the rapid spreading of fake news in social networks. While most of the related work in fake news detection systems has modeled content of the news itself, we propose a complementary approach that takes the network topology and historical user activity into account. 
As the CoViD19 virus spread rapidly around the world in 2020, so has false information regarding various aspects pertaining to 
it\footnote{https://en.wikipedia.org/wiki/Misinformation\_related\_to\_the\_2019-20\_coronavirus\_pandemic}. The need for a spreader detection model for fake news has never been more evident. Thus in this paper, we propose a novel  spreader detection model using an inductive representation learning framework. The model quickly identifies spreaders before the false information penetrates deeper into a densely connected community and infects more nodes. The main contributions of the paper are as follows:
\\
\textbf{1.} We propose a fake news spreader detection framework using the Community Health Assessment model \cite{rath2019evaluating} and interpersonal trust \cite{roy2016trustingness}. To the best of our knowledge, this is the first fake news spreader detection model proposed that relies on features extracted from underlying network structure and historical behavioral data instead of the content.\\
\textbf{2.} We implement our framework using inductive representation learning \cite{hamilton2017inductive} where we sample neighborhood of nodes in a weighted network and aggregate their trust-based features. \\
\textbf{3.} We evaluate our proposed interpersonal trust based framework using multiple real Twitter networks and show that trust based modeling helps us identify false information spreaders with high accuracy, which makes the technique useful for fake news mitigation. \\
\textbf{4.} We further observe that our model's accuracy when detecting false information spreaders is higher than that for true information spreaders. This indicates that people are usually able to reason about true information from analyzing the content, and thus trust in their neighbors is not a very significant factor. However, determining the truth of {\it false information that is plausibly true} from content itself is difficult and hence we have to rely on sources we trust to believe in it or not. This makes nodes that are fake news spreaders and at the same time highly trusted by lots of people in the network, especially dangerous. We acknowledge that not all such {\it uber-spreaders} have ill intentions; some might be just ignorant. They all, nonetheless, have power to spread false information far and wide, with great speed. We believe this phenomenon needs greater study.

The rest of the paper is organized as follows:  We first discuss related work, then describe a motivating example for  spreader detection from a network structure perspective, and summarize past ideas that the proposed research builds upon. We then explain the proposed framework and how we model interpersonal trust with it followed by experimental analysis and finally give our concluding remarks and proposed future work.

\section{Related Work}
In this section we first discuss related works on Graph Neural network architectures. Next, we discuss works related to the application of GNNs to social networks and information dissemination. We then outline other works in the domain of fake news detection, and we finally present works on Inductive representation learning that we build on. 

Graph Neural Networks (GNNs) are powerful neural network models that have received increased attention recently because of their application to non-euclidean spaces such as social networks. Numerous mathematical models for GNNs have been proposed \cite{wu2020comprehensive}. In recent times, there has been research that has leveraged GNNs for complex tasks in social graphs like political perspective prediction and stance detection. In the field of fake news detection, Bian et al. \cite{bian2020rumor} proposed a graph convolution network based model that utilized propagation paths to detect fake news. Researchers have also proposed architectures that integrate ideas from generative adversarial networks to build graph-based detectors for rumor identification \cite{yang2020rumor}. Other studies have demonstrated the merit of attention based graph models in modelling and detecting rumors \cite{yuan2019jointly}, \cite{chen2018call}. Notably, Lu et al. \cite{lu2020gcan} developed a graph-aware attention network that uses user representations and propagation paths taken by a piece of information to predict fake news. Nguyen et al. \cite{Nguyen_2020} recently proposed FANG, an inductive learning framework that uses GNNs for social structure representation and fake news detection. Our work also utilizes inductive representation learning in the form of GraphSage \cite{hamilton2017inductive}, which generates embeddings by sampling and aggregating features. GraphSage generalizes well to unseen and rapidly changing data by dynamically adapting at inference time. 


Characterizing the differences between the spread of false and true news has also served as motivation for our research. In this regard, Vosoughi et al's. \cite{vosoughi2018spread} work on the empirical analysis of the propagation paths taken by false and true news is of interest to us. Jooyeon et al. \cite{kim2019homogeneity} also proposed a bayesian nonparametric model to understand the role of content in diffusion of true and false news and the differences therein.

Unlike most previous works that analyzes content features, our approach uses the underlying social graph structures along with users representations built from their historical data to build an inductive learning based graph neural network to help identify the most prevalent information spreaders.

\section{Motivation and Preliminaries}
\begin{figure}
\centering
\includegraphics[width=\linewidth]{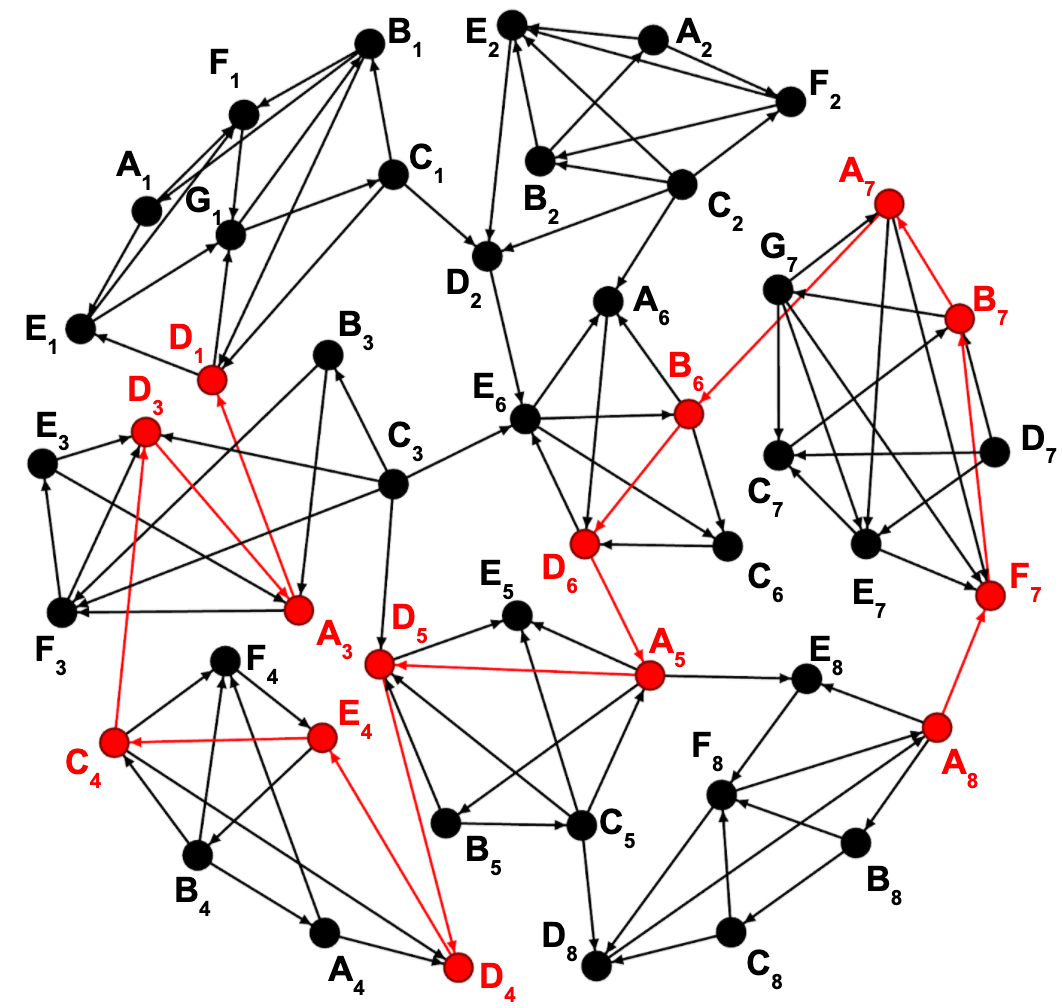}
\caption{Motivating example. Red nodes denote fake news spreaders.}
\label{fig:example}
\end{figure}
To understand the role of network structure in fake news spreader detection, consider the  scenario illustrated in Figure~\ref{fig:example}. The network  contains 8 communities. Subscript of a node denotes the community it belongs to. 
In the context of Twitter, directed edge $B_1 \rightarrow A_1$ represents $B_1$ follows $A_1$. Thus, a tweet flows from $A_1$ to $B_1$. If $B_1$ decides to retweet $A_1$'s tweet, we say that $B_1$ has endorsed $A_1$'s tweet, and that $B_1$ trusts $A_1$. Communities in social networks are \textit{modular} groups, where within-group members are tightly connected, and intra-community trust is higher, compared to trust between members in different communities, who are at best loosely connected. The more $B$ trusts $A$, the higher the chance that $B$ will retweet $A$'s tweet, and thus propagate $A$'s message, whether it is true or false. The figure illustrates the spread of fake news starting from $D_1$ as it spreads across the network through $A_3$ till $A_8$. We consider two scenarios for  spreader detection:\\
\textbf{1. Information reaches neighborhood of a community:} Consider the scenario when a message is propagated by $D_1$, a neighborhood node for community 3. Node $A_3$ is exposed and is likely to spread the information, thus beginning spread of information into a densely connected community. Thus it is important to predict nodes in the boundary of communities that are likely to become information spreaders.\\ 
\textbf{2. Information penetrates the community:} Consider the scenario where $A_3$ decides to propagate a message. Nodes $B_3$, $D_3$ and $E_3$, which are immediate followers of $A_3$ are now exposed to the information. Due to their close proximity, they are vulnerable to believing the endorser. The remaining nodes of the community ($C_3$, $F_3$) are two steps away from $A_3$. Similarly for community 8 when the message has reached node $A_8$, nodes $D_8$ and $F_8$ are one step away and remaining community members ($E_8$, $C_8$, $B_8$) are two steps away. Intuitively, in a closely-knit community structure if one of the nodes decides to spread a piece of information, the likelihood of it spreading quickly within the entire community is very high. Thus it is important to detect nodes within a community that are likely to become information spreaders to protect the health of the entire community.

Above motivation ideas were elaborated in \cite{rathepidemiology}. Next we discuss some concepts used by our proposed model.

\subsection{Community Health Assessment (CHA) model}
Consider the scenario described in Figure~\ref{fig:example}. If a community member believes the information and becomes a spreader, the likelihood of other community members becoming spreaders would be high due to dense connectivity, and hence higher trust, among community members. Using the Community Health Assessment model we propose the ideas of neighbor, boundary and core nodes for every community in a social network. The three types of nodes from  community ($com$) perspective that are affected during the process of news spreading are explained below:
\\
\textit{1. Neighbor nodes} ($\mathcal{N}_{com}$): These nodes are directly connected to at least one node of the community. They are not a part of the community.\\
\textit{2. Boundary nodes} ($\mathcal{B}_{com}$): These are community nodes that are directly connected to at least one neighbor node. It is important to note that only community nodes that have an outgoing edge towards a neighbor node are in $\mathcal{B}_{com}$. \\
 \textit{3. Core nodes} ($\mathcal{C}_{com}$): Community nodes that are only connected to members within the community.  
 
The idea was proposed in \cite{rath2019evaluating} to show how trust plays a more important role in spreading fake news compared to true news. The neighbor, boundary, and core nodes for communities in Figure~\ref{fig:example} are listed in Table~\ref{tab:table1}.

\begin{table}[H]
\caption{Neighbor, boundary and core nodes for communities in Figure~\ref{fig:example}.}
\vspace{.3\baselineskip}
\label{tab:table1}
\resizebox{\linewidth}{!}{%
\begin{tabular}{c|l|l|l}
$com$ & $\mathcal{N}_{com}$ & $\mathcal{B}_{com}$ &  $\mathcal{C}_{com}$  \\
 \hline
1 & $D_2$ & $C_1$ & $A_1$,$B_1$,$E_1$,$D_1$,$F_1$,$G_1$   \\
2 & $A_6$,$E_6$  & $C_2$,$D_2$ & $A_2$,$B_2$,$E_2$,$F_2$ \\
3 & $D_1$,$D_5$,$E_6$ & $A_3$,$C_3$ & $B_3$,$D_3$,$E_3$,$F_3$  \\
4 & $D_3$  & $C_4$ & $A_4$,$B_4$,$D_4$,$E_4$,$F_4$ \\
5 &$D_4$,$D_8$,$E_8$ & $A_5$,$C_5$,$D_5$ & $B_5$,$E_5$   \\
6 & $A_5$ & $D_6$ & $A_6$,$B_6$,$C_6$,$E_6$  \\
7 & $B_6$ & $A_7$ & $B_7$,$C_7$,$D_7$,$E_7$,$F_7$, $G_7$ \\
8 & $F7$ & $A_8$ & $B_8$,$C_8$,$D_8$,$E_8$,$F_8$  
\end{tabular}
}
\end{table}

\subsection{Trustingness and Trustworthiness}
The Trust in Social Media (TSM) algorithm assigns a pair of complementary trust scores to each node in a network called \textit{Trustingness} and \textit{Trustworthiness}.  \textit{Trustingness (ti)} quantifies the propensity of a node to trust its neighbors and \textit{Trustworthiness (tw)} quantifies the willingness of the neighbors to trust the node.
The TSM algorithm takes a user network, i.e., a directed graph $\mathcal{G(\mathcal{V},\mathcal{E})}$, as input together with a specified convergence criteria or a maximum permitted number of iterations. In each iteration for every node in the network, trustingness and trustworthiness are computed using the equations mentioned below:
\begin{align}
ti(v)=&\sum_{\forall x \in out(v)}\left(\frac{w(v,x)}{1+(tw(x))^s}\right) \\
tw(u)=&\sum_{\forall x \in in(u)}\left(\frac{w(x,u)}{1+(ti(x))^s}\right)
\end{align}
where $u, v, x \in \mathcal{V}$ are user nodes, $ti(v)$ and $tw(u)$ are trustingness and trustworthiness scores of $v$ and $u$, respectively, $w(v,x)$ is the weight of edge from $v$ to $x$, $out(v)$ is the set of out-edges of $v$, $in(u)$ is the set of in-edges of $u$, and $s$ is the involvement score of the network. Involvement is basically the potential risk a node takes when creating a link in the network, which is set to a constant empirically. 
The details of the algorithm are excluded due to space constraints and can be found in \cite{roy2016trustingness}.

\subsection{Believability}
\textit{Believability} is an edge score derived from Trustingness and Trustworthiness scores. It quantifies how likely the receiver of a message is to believe its sender. Believability for a directed edge is naturally computed as a function of the trustworthiness of the sender and the trustingness of the receiver.
So, the believability score is supposed to be proportional to the two values above, which can be jointly determined and computed as follows:
\begin{equation}\label{eq:believability}
bel_{uv} = tw(u) * ti(v)
\end{equation}
The idea has been applied in \cite{rath2018utilizing} where an RNN model was proposed to identify rumor spreaders in Twitter networks.

\section{Proposed Approach}
\textbf{Problem Formulation:} Given a directed social network $\mathcal{G(\mathcal{V},\mathcal{E})}$ comprising disjoint modular communities ($\phi$), with each community ($com \in \phi$) having well-defined neighbor nodes ($\mathcal{N}_{com}$), boundary nodes ($\mathcal{B}_{com}$) and core nodes ($\mathcal{C}_{com}$). Aggregating topology-based ($top$) and activity-based ($act$) trust properties from nodes sampled from depth $K$ (where $Nbr_{K=1}(b) \subseteq \mathcal{N}_{com}$), we want to predict boundary nodes $b$ that are most likely to become information spreaders ($b_{sp}$). Similarly, we aggregate nodes sampled from depth $K$ (where $Nbr_{K=1}(c) \subseteq \mathcal{B}_{com}$) to predict core nodes $c$  that are most likely to become information spreaders ($c_{sp}$).

\begin{figure*}
\begin{subfigure}{.33\linewidth}
\centering
\includegraphics[width=2.4in]{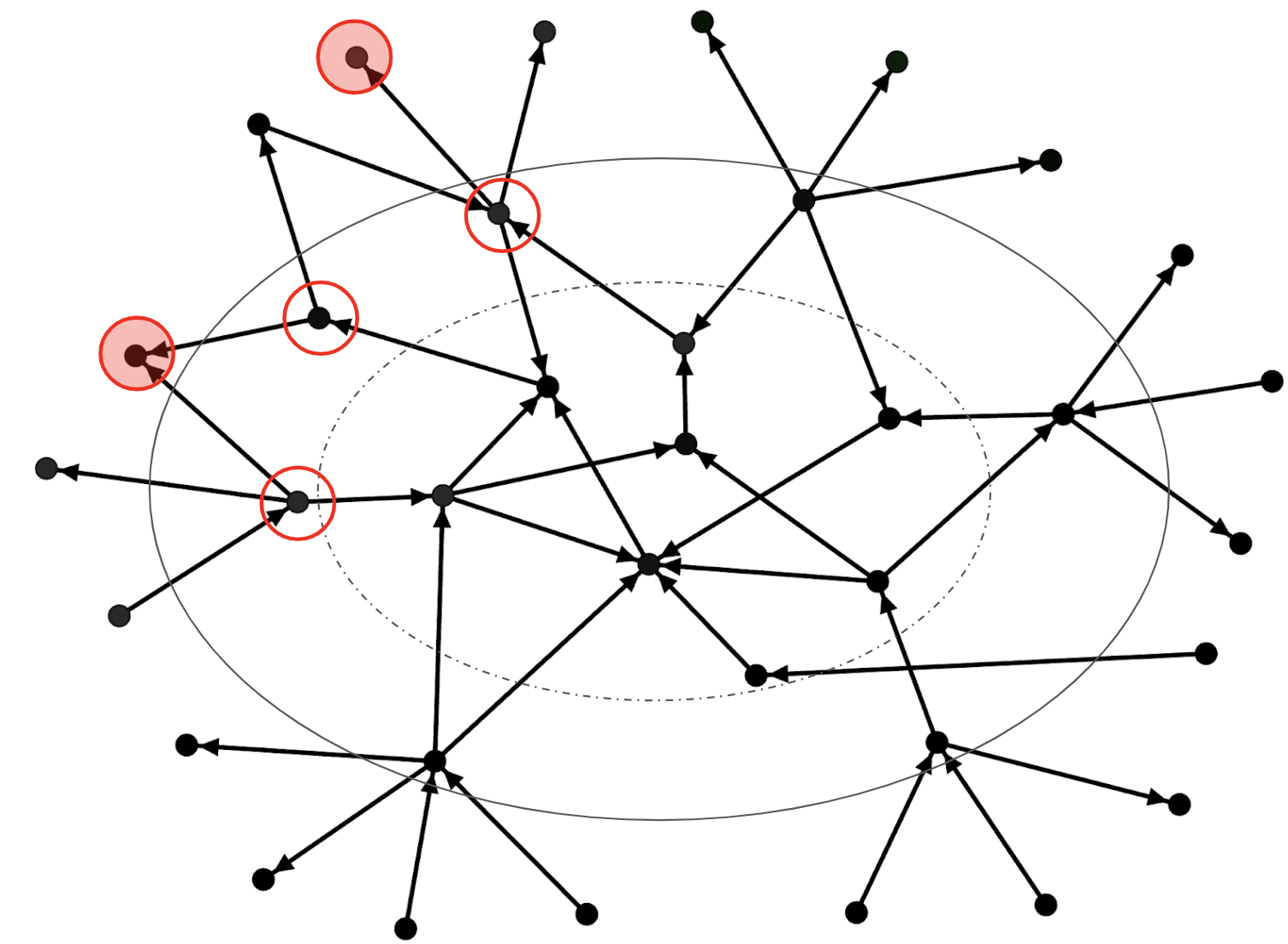}
\caption{Information reaches $\mathcal{N}_{com}$}\label{fig:mouse}
\end{subfigure}
\begin{subfigure}{.33\linewidth}
\includegraphics[width=2.4in]{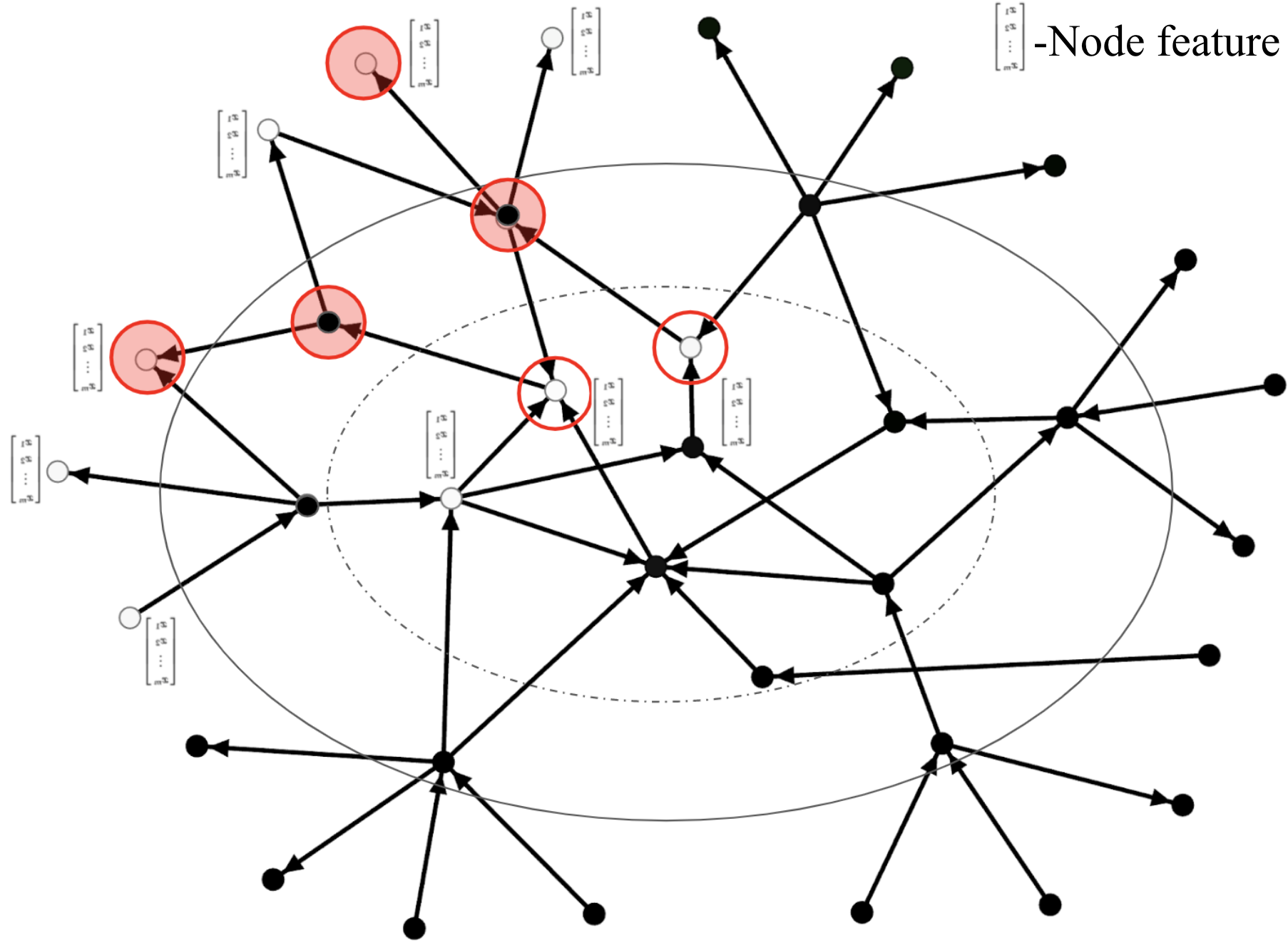}
\caption{Information reaches $\mathcal{B}_{com}$}\label{fig:gull}
\end{subfigure}
\begin{subfigure}{.3\linewidth}
\includegraphics[width=2.4in]{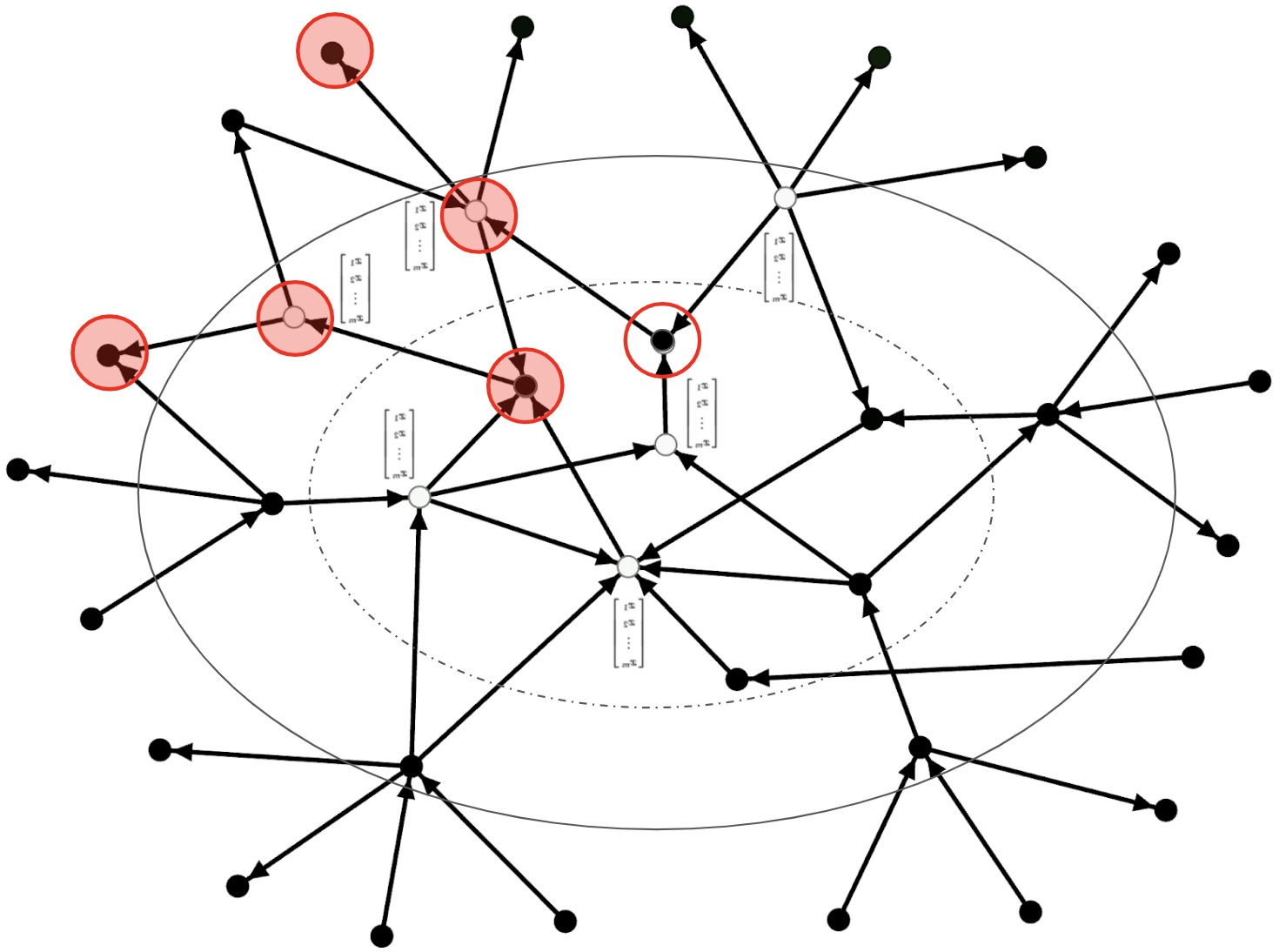}
\caption{Information reaches $\mathcal{C}_{com}$}\label{fig:gull}
\end{subfigure}
\caption{Inductive representation learning model for detection of fake news spreaders.}
\label{fig:model}
\end{figure*}

{\bf Inductive Representation Learning:} As fake news spreads rapidly, network structure around the spreaders also evolves quickly. Thus, it is important to have a scalable model that is able to quickly learn meaningful representations for newly seen (i.e. exposed) nodes without relying on the complete network structure. Most graph representation learning techniques, however, employ a \textit{transductive} approach to learning node representations which optimizes the embeddings for nodes based on the entire graph structure. We employ an \textit{inductive} approach inspired from  GraphSAGE \cite{hamilton2017inductive} to generate embeddings for the nodes as the information spreading network gradually evolves. It learns an aggregator function that generalizes to unseen node structures which could become potential information spreaders. The idea is to simultaneously learn the topological structure and node features from the neighborhood ($Nbr$) nodes, by training a set of aggregator functions instead of individual node embeddings. Using an inductive representation learning model we learn features of the exposed population (i.e. followers of the spreaders) by aggregating trust-based features from their neighborhood nodes. Figure~\ref{fig:model} shows how we model the proposed approach with community perspective. Nodes outside the solid oval represent $\mathcal{N}_{com}$, between solid and dotted oval represents $\mathcal{B}_{com}$ and within the dotted oval  represents $\mathcal{C}_{com}$. (a) shows that false information spread has reached the two neighbor nodes (highlighted in red). Three boundary nodes (circled in red) are exposed to the information. In (b) we learn  representations for the exposed boundary nodes by aggregating features of their local neighborhood structure (denoted by white nodes). Two out of the three  boundary nodes that become spreaders are highlighted and the exposed core nodes are circled. Similarly, in (c) we learn  representations for the exposed core nodes by aggregating their local neighborhood features. One core node becomes a spreader and the community is now vulnerable to fake news spreading.

The proposed framework is explained as follows: First we generate a weighted information spreading network based on interpersonal trust. We then sample neighborhood with a probability proportional to the trust based edge weights. For the sampled neighborhood we aggregate their feature representations. Finally we explain the loss function used to learn  parameters of the model.
\subsection{Generating weighted graph}
Graph of the information spreading network has edge weights that quantify the likelihood of trust formation between senders and receivers. Once we compute these edge scores using techniques mentioned in Table~\ref{tab:modelling}, we normalize weights for all out-edges connecting the boundary node.  
\begin{equation}
\hat{w}_{bx}=\frac{bel_{bx}}{\sum_{\forall x \in out(b)}bel_{bx}} \\
\label{eq:w}
\end{equation}
Similarly we normalize weights for all in-edges connecting the boundary node.

\subsection{Sampling neighborhood}
Instead of sampling neighborhood as a uniform distribution, we sample a subset of neighbors proportional to the weights of the edges connecting them. Sampling is done recursively till depth $K$. The idea is to learn features from neighbors proportional to the level of inter-personal trust. Algorithm~\ref{algorithm1} explains the sampling strategy.

\begin{algorithm}
  $\textbf{Input:} \quad \text{$\mathcal{G(\mathcal{V},\mathcal{E})}$: Information spreading network}$,\\ $K$: Sampling depth, $\mathcal{B}_{com}$: Boundary nodes of community.\\
  $\textbf{Output:}$ $Nbr_{K}(b)$: Sampled neighborhood for $b$ till depth $K$.
  \\
  $ \phi \gets \text{Disjoint modular communities in $\mathcal{G}$}$\;
   \For{each $com \in \phi$}
  {
     \For{each $b \in \mathcal{B}_{com}$}
     {
     $Nbr_{0}(b) \gets \{b\}$\\
     \For{ $k = 1 \dots K$}
         {
         $Nbr_{k}(b) \gets Nbr_{k-1}(b) \cup{SA_{k}(b)}_{ Eq ~\ref{eq:w}}$
         }
     }
  }
  \caption{Sample neighborhood ($SA$)}
  \label{algorithm1}
\end{algorithm}

 \subsection{Aggregating features}
After sampling neighborhood as an unordered set, we aggregate the embeddings of sampled nodes till depth $K$ recursively for each boundary node. The intuition is that at each depth, the boundary nodes incrementally learn trust-based features from the sampled neighborhood. Three aggregation architectures namely mean, LSTM and pooling explained in \cite{hamilton2017inductive} can be used. For simplicity, we only apply the mean aggregator, which takes the mean of representations $h^{k-1}_u$  where $u \in Nbr_{k-1}(b)$. The aggregator is represented below:
\begin{equation}
h^{k}_b \gets \sigma(W^{k}_{b}.Mean(\{h^{k-1}_b\} \cup \{h^{k-1}_{u(\forall u \in Nbr(b))})\}) \\
\label{eq:s}
\end{equation}
Algorithm~\ref{algorithm2} explains the aggregation strategy.

\begin{algorithm}
  $\textbf{Input:} \quad \text{$\mathcal{G(\mathcal{V},\mathcal{E})}$: Information spreading network}$,\\ $K$: Sampling depth, $\mathcal{B}_{com}$: Boundary nodes of community, $x_{v(\forall v \in \mathcal{V})}$: Node features.\\
  $\textbf{Output:}$ $z^{k}_{b}$: Embedding vector for $b$.
  \\
  $ \phi \gets \text{Disjoint communities in $\mathcal{G}$}$\;
   \For{each $com \in{} \phi$}
  {
     \For{each $b \in{} \mathcal{B}_{com}$}
     {
     $h^{0}_{b}  \gets x_{b}$\\
     \For{ $k = 1 \dots K$}
         {
            $h^{k}_{Nbr(b)} \gets GE_{k}(h^{k-1}_{u(\forall u \in Nbr(b))})$\\
            $h^{k}_{b} \gets \sigma (W^{k}_{b}.Concat(h^{k-1}_{b},    h^{k}_{Nbr(b)}))_{ Eq. ~\ref{eq:s}}$
         }
         $h^{k}_{b} \gets h^{k}_{b}/ ||h^{k}_{b}||_{2}$
     }
      $z^{k}_{b} \gets h^{k}_{b}$
  }
  
  \caption{Aggregate features ($GE$)}
  \label{algorithm2}
\end{algorithm}

\subsection{Learning parameters}
The weight matrices in Algorithm~\ref{algorithm2} are tuned using stochastic gradient descent on a loss function in order to learn the parameters. We train the model to minimize cross-entropy.
\begin{equation}
Loss(\hat{y}, y) = - \sum_{\forall b \in \mathcal{B}_{com}} \sum_{i \in \{b_{Sp},b_{\bar{Sp}}\}}y_{i}log \hat{y}_{i}\\
\label{eq:param_learn}
\end{equation}
The loss function is modeled to predict whether the boundary node is an information spreader ($b_{Sp}$) or a non-spreader ($b_{\bar{Sp}}$). $y$ represents the actual class (2-dimensional multinomial distribution of [1,0] for spreader and [0,1] for non-spreader) and $\hat{y}$ represents the predicted class.

We extend the model for $\mathcal{C}_{com}$ to identify the core node spreaders ($c_{Sp}$) and non-spreaders ($c_{\bar{Sp}}$). Considering boundary nodes have denser neighborhood compared to core nodes, we later analyze whether the proposed model is more sensitive to density of neighborhood structure or the aggregated features. The implementation code is made publicly available\footnote{https://github.com/BhavtoshRath/Proactive\_Spreader\_Detection}.
\subsection{Modeling interpersonal trust}
As explained in the preliminaries section, interpersonal  trust has been applied successfully in the past to model spreading of fake news. Thus we model our node representation learning problem using interpersonal trust to predict whether a node is a spreader or not. We first apply a non-uniform neighborhood sampling strategy using weighted graph (where edge weights quantify the likelihood of trust formation). We then aggregate two trust features: 1) The likelihood of \textit{trusting others} and 2) The likelihood of being \textit{trusted by others}. We use two kinds of interpersonal-trust: Topology-based ($top$) computed from the social network topology and  Activity-based ($act$) computed using timeline activity data collected for every node using Twitter API. We use trustingness ($ti(x)$) and trustworthiness ($tw(x)$) scores of node $x$ obtained from TSM as proxy for topology-based trust features and  the fraction of timeline statuses of $x$ that are retweets ($RT_x$) denoted by $\sum_{\forall i \in t}\{1$ if $ i=RT_x$ else $0\}/n(t)$ and average number of times $x$'s tweets are retweeted ($n(RT_{x})$) denoted by $\sum_{\forall i \in t}{{{i_{n(RT_x)}}}/n(t)}$ as activity-based trust features ($t$ represents most recent tweets posted on $x$'s timeline\footnote{Due to time restrictions we collected only 10 most recent tweets for every node in the network.}). For an edge from $x$ to $v$, the topology-based edge weight is the believability score ($bel_{xv}$) and activity-based edge weight is the number of times  $x$ is retweeted by $v$ ($RT_{xv}$). Trust-based sampling and aggregation strategy is summarized in Table~\ref{tab:modelling}.

\begin{table}[h]
\caption{Trust based strategy for sampling and aggregating.}
\vspace{.3\baselineskip}
\label{tab:modelling}
\resizebox{\linewidth}{!}{%
\begin{tabular}{c|c|c|c}
& & \textbf{Topology ($top$)} & \textbf{Activity ($act$)}    \\
 \hline
$Sample$ & $w_{xv}$ & $bel_{xv}$ & $RT_{xv}$  \\
\hline
\multirow{2}{*}{$Aggregate$} 
&   $trusting$ $others$  & $ti(x)$ & 
        $\frac{\sum_{\forall i \in t}  
        \begin{cases}
        1 & \text{if $i=RT_x$}\\
        0 & \text{otherwise.}
    \end{cases}   }{n(t)}$
    \\
& $trusted$ $by$ $others$ & $tw(x)$  & $\frac{\sum_{\forall i \in t}{{i_{n(RT_x)}}}}{n(t)}$\\
\end{tabular}
}
\end{table}

\section{Experiments and Results}

\subsection{Ground truth and data collection}
We evaluate our proposed model using real world Twitter datasets. We obtained the ground truth of false information and the refuting true information from \textit{altnews.in}, a popular fact checking website. The source tweet related to the information was obtained directly as a tweet embedded in the website or through a keyword based search on Twitter. From the source tweet we generated the source tweeter and the retweeters (proxy for spreaders), follower-following network of the spreaders (proxy for network) and the timeline data for all nodes in the network (to generate trust-based features) using the Twitter API. Besides evaluating our model on false information (F) and the refuting true information (T) networks separately, we also evaluated on network obtained by combining them (F $\cup$ T). Metadata for the network dataset aggregated for all news events is summarized in Table~\ref{tab:dataset}.

\begin{table}[H]
    \caption{Network dataset statistics.}
    \vspace{.3\baselineskip}
    \label{tab:dataset}
    \resizebox{\linewidth}{!}{%
    \begin{tabular}{l|c|c|c} 
    \textbf{} & \textbf{F} & \textbf{T} & \textbf{F $\cup$ T} \\
    \hline
    No. of nodes  & 1,709,246 & 1,161,607 & 2,554,061 \\
    No. of edges  & 3,770,532 & 2,086,672 & 5,857,205 \\
    No. of  spreaders   & 2,246 & 643 & 2,862 \\
    No. of communities & 58 & 39  & 52 \\
    No. of nodes in $\mathcal{N}$    & 209,311 & 94,884 & 276,567 \\
    No. of spreaders in $\mathcal{N}$  & 19,403 & 5,350 & 22,868 \\
    No. of nodes in $\mathcal{B}$    &217,373  & 136,350 & 345,312 \\
    No. of spreaders in $\mathcal{B}$   & 2,152 & 611 & 2,738 \\
    No. of nodes in $\mathcal{C}$    &1,278,885  & 862,778 & 1,893,493 \\
    No. of spreaders in $\mathcal{C}$    & 94 & 31 & 98 \\
    \end{tabular}
    }
\end{table}

\begin{table*}
        \caption{Results comparison of different models for boundary node spreader prediction.}
    \label{tab:boundary_pred}
        \resizebox{\textwidth}{!}{%
    \begin{tabular}{l|c|c|c|c|c|c|c|c|c|c|c|c} 
\multirow{2}{0.5cm}
    \textbf{} &
    \multicolumn{4}{c|}{\textbf{F}} & 
    \multicolumn{4}{c|}{\textbf{T}} &
    \multicolumn{4}{c}{\textbf{F $\cup$ T}}\\\cline{2-13}
    \textbf{} & \textbf{Accu.} & \textbf{Prec.} & \textbf{Rec.} & 
    \textbf{F1} & \textbf{Accu.} & \textbf{Prec.} & \textbf{Rec.} & 
    \textbf{F1} & \textbf{Accu.} & \textbf{Prec.} & \textbf{Rec.} & 
    \textbf{F1} \\
    \hline
    $Trusting$ $others$ & 0.58 & 0.612 & 0.329 & 0.396 & 0.615 & 0.697 & 0.450 & 0.519  & 0.510 & 0.522 & 0.888 & 0.603 \\
    $Trusted$ $by$ $others$ & 0.608 & 0.631 & 0.384 & 0.455 & 0.646 & 0.713 & 0.500 & 0.585 & 0.518 & 0.513 & 0.916 & 0.638\\
    $Interpolation$ & 0.622 & 0.635 & 0.426 & 0.498 & 0.661 & 0.768 & 0.496 & 0.588 & 0.524 & 0.526 & 0.846 & 0.611\\
    $LINE$ & 0.709 & 0.784 & 0.593 & 0.669 & 0.692 & 0.763 & 0.567 & 0.647 & 0.589 & 0.602 & 0.517 & 0.554 \\
    $GCN_{top}$ & 0.839 & 0.887 & 0.784 & 0.832 & 0.775 & 0.921 & 0.595 & 0.723 & 0.592 & 0.649 & 0.646 & 0.647\\
    $GCN_{act}$ & 0.807 & 0.849 & 0.750 & 0.796 & 0.740 & 0.835 & 0.591 & 0.693 & 0.576 & 0.640 & 0.612 & 0.626\\
    $SA_{rand}GE_{top}$ & 0.870 & 0.879 & 0.862 & 0.866 & 0.776 & 0.858 & 0.667 & 0.748 & 0.599 & 0.605 & 0.570 & 0.583\\
    $SA_{rand}GE_{act}$ & 0.777 & 0.845 & 0.689 & 0.754 & 0.728 & 0.814 & 0.612 & 0.688 & 0.566 & 0.572 & 0.539 & 0.547\\
    $SA_{top}GE_{top}$ & \textbf{0.937} & \textbf{0.918} & \textbf{0.965} & \textbf{0.939} & \textbf{0.834} & \textbf{0.927} & \textbf{0.732} & \textbf{0.815} & \textbf{0.616} & \textbf{0.630} & \textbf{0.561} & \textbf{0.592}\\
    $SA_{top}GE_{act}$ & 0.912 & 0.899 & 0.935 & 0.915 & 0.800 & 0.884 & 0.699 & 0.777 & 0.584 &0.601  & 0.504 & 0.545\\
    $SA_{act}GE_{top}$ & 0.838 & 0.854 & 0.816 & 0.833 & 0.763 & 0.817 & 0.686 & 0.743 & 0.582 & 0.589 & 0.542 & 0.559\\
    $SA_{act}GE_{act}$ & 0.804 & 0.853 & 0.737 & 0.786 & 0.735 & 0.800 & 0.634 & 0.706 & 0.561 & 0.570 & 0.542 & 0.539\\
   \end{tabular}
    }
\end{table*}

  \begin{figure*}
\centering
\includegraphics[width=\linewidth]{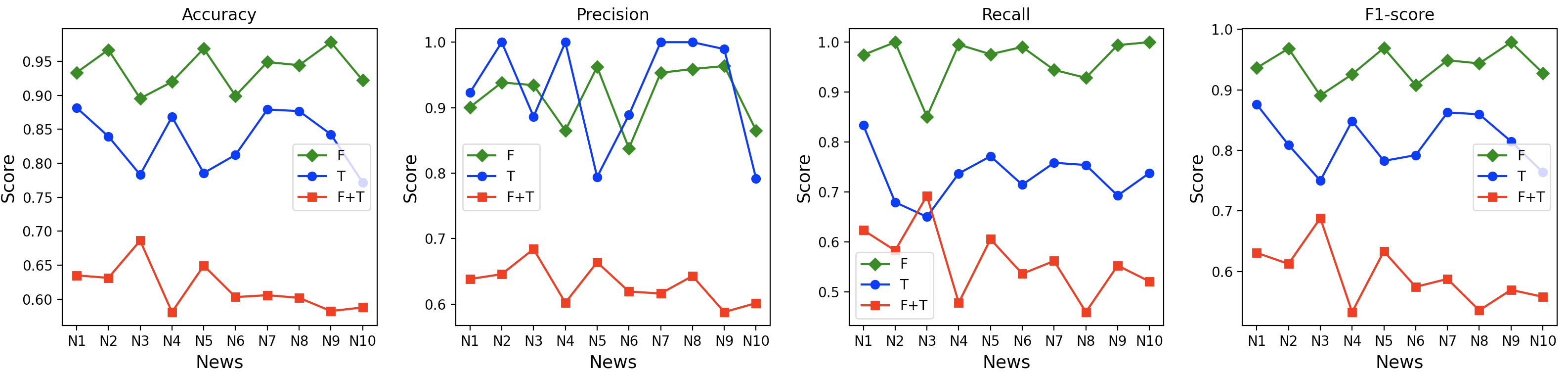}
\caption{Metric performance of boundary node prediction by $SA_{top}GE_{top}$ model for news N1 to N10.}
\label{fig:b_6}
\end{figure*}

\subsection{Settings and Protocols}
We obtained the topology-based measures by running TSM algorithm on the network to obtained $ti$, $tw$  for all nodes and $bel$ for all edges. We used the generic settings for TSM parameters (number of iterations = 100, involvement score = 0.391) by refering to \cite{roy2016trustingness}. We found the  disjoint modular communities using Louvain community detection algorithm \cite{blondel2008fast} and identified the neighbor, boundary and core nodes for every community using Community Health Assessment model. We then generated the activity-based measures from timeline data of the nodes. The embeddings are generated using the forward propagation method shown in Algorithm~\ref{algorithm2}, assuming that the model parameters are learnt using Equation~\ref{eq:param_learn}. Due to class imbalance we undersample the majority class to obtain balanced spreader and non-spreader class distribution. The size of hidden units is set to 128 and the learning rate is set to 0.001. We used rectified linear units as the non-linear activation function. The batch size was adjusted for optimal performance depending on the size of training dataset. Due to the heavy-tailed nature of degree distributions of edges in social networks we downsample before modeling, which ensured that the neighborhood information is stored in dense adjaceny lists. This drastically reduces our run time, which is ideal for early detection of spreaders. We also set sampling depth $K$=1 because the network constitutes only immediate follower-following nodes of the spreaders. We compared results for the following models, including baselines: \\
    1) $Trusting$ $others$: Intuitively, users with high likelihood to trust others tend to be spreaders of false information. This model learns a threshold based on correlation between 'trusting others' features (both topology- and activity- based) and user ground truth. \\
    2) $Trusted$ $by$ $others$: Intuitively, users with high likelihood to be trusted by others tend to be spreaders of false information. Like the previous model, this model learns a threshold based on correlation between 'trusted by others' features (both topology- and activity- based) and user ground truth.\\ 
    3) $Interpolation$: This model linearly combines 'trusting others' and 'trusted by others' features to find an optimal threshold.\\
    4) $LINE$: This model applies LINE \cite{tang2015line} which serves as transductive learning baseline.\\
    5) $GCN_{top}$: This model implements graph convolutional networks \cite{kipf2016semi} based transductive learning model that aggregates topology features from neighborhood.\\
    6) $GCN_{act}$: This is the graph convolutional networks based model that aggregates activity features from neighborhood.\\
    7) $SA_{rand}GE_{top}$: This model applies the inductive learning by sampling neighborhood considered as uniform distribution and aggregating only topology based features.\\
    8) $SA_{rand}GE_{act}$: This model applies the inductive learning by sampling neighborhood considered as uniform distribution and aggregating only activity based features.\\
    9) $SA_{top}GE_{top}$: Instead of random sampling, we sample on the believability ($bel$) weighted network and aggregate their topology based features.\\
    10) $SA_{top}GE_{act}$: Sampling approach is identical to 11) but we aggregate neighborhood's activity based features.\\
    12) $SA_{act}GE_{top}$: We sample neighborhood non-uniformly on the retweet count ($RT$) weighted network and aggregate their topology based features.\\
    13) $SA_{act}GE_{act}$: Sampling approach is identical to 14) but we aggregate neighborhood's activity based features.\\

\begin{table*}
        \caption{Results comparison of different models for core node spreader prediction.}
    \label{tab:core_pred}
        \resizebox{\textwidth}{!}{%
    \begin{tabular}{l|c|c|c|c|c|c|c|c|c|c|c|c} 
\multirow{2}{0.5cm}
    \textbf{} &
    \multicolumn{4}{c|}{\textbf{F}} & 
    \multicolumn{4}{c|}{\textbf{T}} &
    \multicolumn{4}{c}{\textbf{F $\cup$ T}}\\\cline{2-13}
    \textbf{} & \textbf{Accu.} & \textbf{Prec.} & \textbf{Rec.} & 
    \textbf{F1} & \textbf{Accu.} & \textbf{Prec.} & \textbf{Rec.} & 
    \textbf{F1} & \textbf{Accu.} & \textbf{Prec.} & \textbf{Rec.} & 
    \textbf{F1} \\
    \hline
    $Trusting$ $others$ & 0.553 & 0.643 & 0.298 & 0.388 & 0.569 & 0.585 & 0.338 &  0.414 & 0.521 & 0.511 & 0.95 & 0.659\\
    $Trusted$ $by$ $others$ & 0.569 & 0.628 & 0.411 & 0.481 & 0.614 & 0.694 & 0.503 & 0.508 & 0.540 & 0.523 & 0.952 & 0.673\\
    $Interpolation$ & 0.609 & 0.730  & 0.400 & 0.492 &   0.640 & 0.681 & 0.438 & 0.521 & 0.550 & 0.548 & 0.764 & 0.608\\
    $LINE$ & 0.721 & 0.821 & 0.625 & 0.681 & 0.672 & 0.870 & 0.467 & 0.579 & 0.577 & 0.572 & 0.676 & 0.602\\
    $GCN_{top}$ & 0.755 & 0.972 & 0.524 & 0.681 & 0.739 & 0.698 & 0.839 & 0.762 & 0.683 & 0.731 & 0.537 & 0.619\\
    $GCN_{act}$ & 0.731 & 0.741 & 0.705 & 0.722 & 0.701 & 0.735 & 0.641 & 0.684 & 0.657 & 0.691 & 0.561 & 0.619\\
    $SA_{rand}GE_{top}$ & 0.842 & 0.900 &0.802  & 0.838 & 0.726 & 0.880 & 0.574 & 0.664 & 0.656 & 0.651 & 0.707 & 0.665\\
    $SA_{rand}GE_{act}$ & 0.798 & 0.893 & 0.700 & 0.764 & 0.658 & 0.742 & 0.448 & 0.523 &0.597  & 0.631 & 0.512 & 0.548\\
    $SA_{top}GE_{top}$ & \textbf{0.916} & \textbf{0.940} & \textbf{0.892} & \textbf{0.912}  & \textbf{0.836} & \textbf{0.895} & \textbf{0.787} & \textbf{0.825} & \textbf{0.734} & \textbf{0.725} & \textbf{0.823} & \textbf{0.750}\\
    $SA_{top}GE_{act}$ & 0.891 & 0.929 & 0.849 & 0.884 & 0.800 & 0.931 & 0.684 & 0.769 & 0.685 & 0.703 & 0.677 & 0.682\\
    $SA_{act}GE_{top}$ & 0.868 & 0.941 & 0.788 & 0.854 & 0.771 & 0.962 & 0.598 & 0.712 & 0.648 & 0.688 & 0.651 & 0.641\\
    $SA_{act}GE_{act}$ &0.846  & 0.847 & 0.858 & 0.846 & 0.707 & 0.827 & 0.581 & 0.661 & 0.619 & 0.694 & 0.522 & 0.567\\
   \end{tabular}
    }
\end{table*}

\begin{figure*}
\centering
\includegraphics[width=\linewidth]{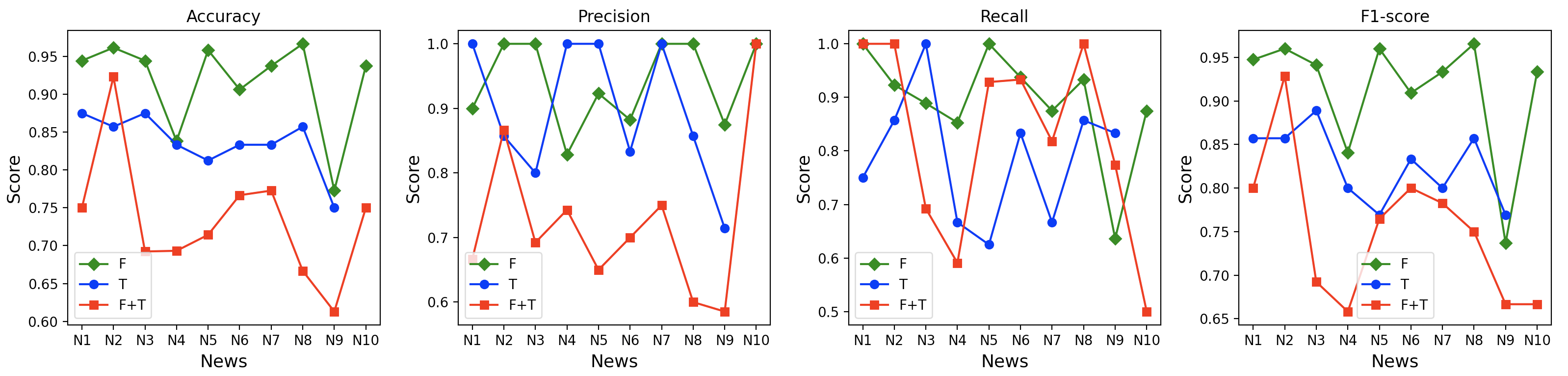}
\caption{Metric performance of core node prediction by $SA_{top}GE_{top}$ model for news events N1 to N10.}
\label{fig:c_6}
\end{figure*}

Baseline models 1) - 3) are inspired from \cite{rath2018utilizing} that considers features based on trust. Baseline model 4) considers features based on network structure only. Proposed models 5) - 13) integrate both neighborhood structure and node features. We analyze the best combination of sampling and aggregating strategy that predicts spreader node with highest accuracy. 
For evaluation we did a 80-10-10 train-validation-test split of the dataset. We used 5-fold cross validation and four common metrics: Accuracy, Precision, Recall and F1 score. 
We only show results for the spreader class.

\subsection{Results and Analysis}
We evaluated our proposed model on 10 debunked news events. For each news event we obtained three types of networks: network for the false information (F), for the true information (T) refuting it and the network obtained by combining them (F $\cup$ T). Thus we ran our models on 30 large-scale networks.\\
\textbf{Boundary node analysis (Less dense \textit{Nbr}):} Table~\ref{tab:boundary_pred} summarizes results for the boundary node prediction aggregated for all news. The results show that F  performs better than T  on almost every metrics while F $\cup$ T  performs poorly. The poor performance of F $\cup$ T networks could be attributed to the fact that there is minimal overlap of nodes in F and T networks (12\%) which causes the F $\cup$ T networks to have sparser communities. Also false and true information spreaders are together considered as spreader class which could be affecting the model performance. While comparing the baseline models, $Trusted$ $by$ $others$ model performs better than the $Trusting$ $others$ model with an improvement in accuracy of 4.8\%, 5\% and 1.5\% for F, T and F $\cup$ T networks respectively. $Interpolation $ model shows a further improvement of 2.3\%, 2.3\% and 1.1\% for F, T and F $\cup$ T networks respectively over $trustingness$ model. $LINE$ and $GCN$ baselines  show significant improvement on all metrics for F networks compared to T or F $\cup$ T networks. We see further substantial increase in performance for each type of network using inductive learning models. Comparing the two random sampler models (i.e. $SA_{rand}GE_{top}$, $SA_{rand}GE_{act}$) we see that topology-based features of the neighborhood perform better than activity-based features. Similar trend is observed for topology-based sampler models (i.e. $SA_{top}GE_{top}$, $SA_{top}GE_{act}$) where model using topology-based aggregator performs better than activity-based aggregator. Same is the case for activity-based sampler models (i.e. $SA_{act}GE_{top}$, $SA_{act}GE_{act}$). Integrating $top$ and $act$ does not show any significant improvement over $top$ only models.
Thus we can conclude that interpersonal trust based modeling in the inductive learning framework is able to  predict false information spreaders better than true information spreaders. We also observe that topology-based sampling and aggregating strategies perform better than activity-based strategies. The low performance of activity-based strategies could be attributed to the fact that many Twitter users are either inactive users or users with strict privacy settings whose timeline data could not be retrieved. Also recent 10 activities on a user's timeline might be insufficient data to capture activity-based trust dynamics. For each type of network, we observe that $SA_{top}GE_{top}$ model performs the best, with F having accuracy of 93.3\%, which is higher than 12.3\% and 52.1\% over T and F $\cup$ T networks respectively.  Figure~\ref{fig:b_6} shows the  performance metrics of this model for the 10 news events (N1-N10). We observe a clear distinction in performance, with F networks performing better than T, which in turn is better than F $\cup$ T. An interesting observation is the high precision values for T. This is because the percentage of predicted spreaders which are non-spreaders tends to be lower for T network than for F network.\\
\textbf{Core node analysis (More dense \textit{Nbr}):} Table~\ref{tab:core_pred} summarizes results of the model for predicting core nodes aggregated for all news. The overall performance trend is identical to the results shown for boundary nodes in Table~\ref{tab:boundary_pred}. Among the baseline models, $Interpolation$ model performs better than $Trusted$ $by$ $others$ and $Trusting$ $others$ models. $LINE$ and $GCN$ based models show  significant improvement over trust feature baselines on all metrics. Among inductive learning models, topology-based trust modeling shows better performance than activity-based trust modeling. Also F networks perform better than T networks, which in turn perform better than F $\cup$ T networks. Among random sampler models, $SA_{rand}GE_{top}$ has the highest accuracy of 84.2\%, 72.6\% and 65.6\% for F, T and F $\cup$ T networks respectively. Among topology-based sampler models $SA_{top}GE_{top}$ performs better over $SA_{top}GE_{act}$ with an increase in accuracy of  2.8\%, 4.5\% and 7.1\% for F, T and F $\cup$ T networks respectively. Activity-based sampler models also show identical trend with $SA_{act}GE_{top}$ performing better than $SA_{act}GE_{act}$ with an increase in accuracy of 2.6\%, 9\% and 4.6\% for F, T and F $\cup$ T networks respectively. Among all models $SA_{top}GE_{top}$ shows the best overall performance.  Figure~\ref{fig:c_6} shows the metric performance of this model for the 10 news events. True information network for N10 is excluded from analysis as it did not have sufficient spreaders to train our model on. A clear observation is that the metric performance for the three types of networks is not as distinct as in  Figure~\ref{fig:b_6}. Even though the number of core nodes is much higher than boundary nodes, the number of core spreaders is much smaller than boundary node spreaders. Thus the model fails to learn meaningful representations for core nodes due to smaller training dataset.\\
\textbf{Summary:} Comparing the prediction performance of core and boundary spreaders we can conclude that our model's performance is more sensitive to aggregated features and training dataset size compared to density of neighborhood. 


\section{Conclusions and future work}
In this paper we proposed a novel fake news spreader detection model for communities using inductive representation learning and community health assessment. Using interpersonal trust based properties we could identify spreaders with high accuracy, and also showed that the proposed model identifies false information spreaders more accurately than true information spreaders. 
The key hypothesis we tested is that interpersonal trust plays a significantly more important role in identifying false information spreaders than true information spreaders. 
Identified false information spreaders can thus be quarantined and true news spreaders can be promoted, thus serving as an effective mitigation strategy. Experimental analysis on Twitter data  showed that topology-based modeling yields better results compared to activity-based modeling. The proposed research can be used to identify people who are likely to become spreaders in real-time due to its ability to adapt to rapidly evolving information spreading networks. As part of future work we want to test our  model on higher volume of user timeline activity which would give a better picture of the effectiveness of the activity-based approach. 
We would also want to take into consideration the presence of bots. 
We would also want to extend the network further in order to sample neighborhood from greater sampling depths. \\
%
%

\bibliographystyle{IEEEtran}
\end{document}